\documentclass[%
reprint,
superscriptaddress,
 aps,
prstab,
floatfix
]{revtex4-2}

\usepackage{graphicx}
  \setkeys{Gin}{width=\linewidth}
\usepackage{dcolumn}
\usepackage{bm}

\usepackage{amsmath,amsfonts,amssymb}
\usepackage{amsbsy}

\usepackage{hyperref}
\hypersetup{
    colorlinks=true,       
    linkcolor=blue,          
    citecolor=blue,        
    filecolor=blue,         
    urlcolor=blue        
}

\newcommand{\orcidicon}[1]{\href{https://orcid.org/#1}{\includegraphics[height=\fontcharht\font`\B,keepaspectratio]{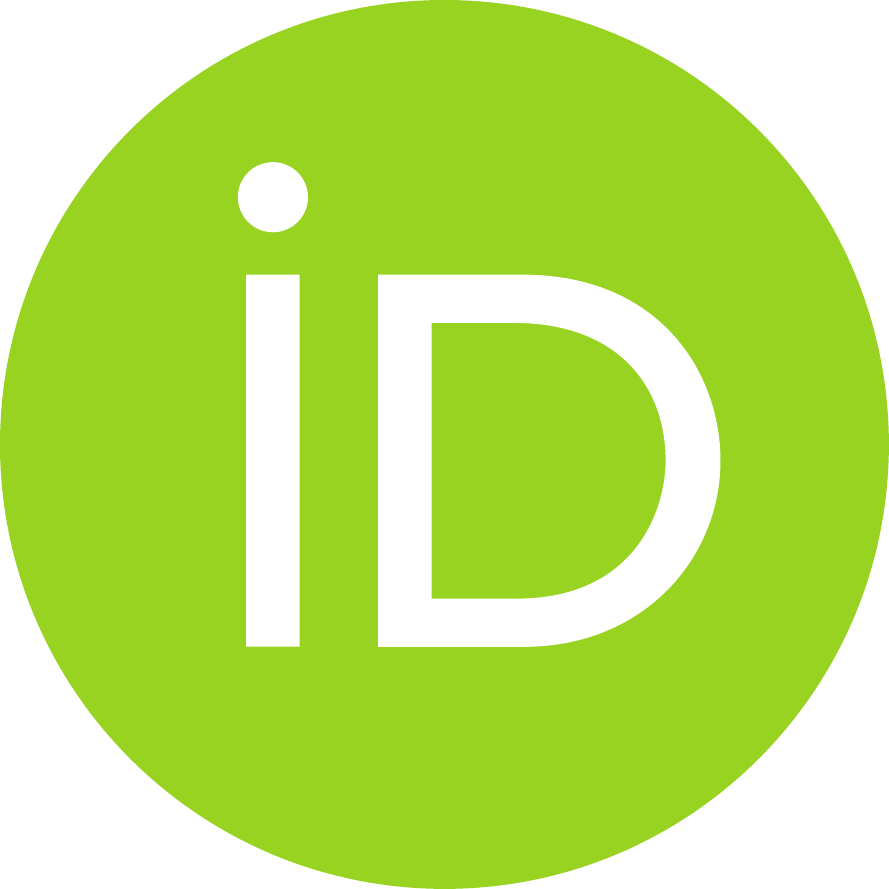}}}

\begin{document}


\title{Na-intercalation compounds and Na-ion batteries}

\author{Wei-Bang Li\,\orcidicon{0000-0002-8319-3316}}
\email[E-mail: ]{weibang1108@gmail.com}
\affiliation{Department of Physics, National Cheng Kung University, Tainan 70101, Taiwan}

\author{Yu-Ming Wang\,\orcidicon{0000-0001-8866-065X}}
\email[E-mail: ]{wu0h96180@gmail.com}
\affiliation{Department of Physics, National Cheng Kung University, Tainan 70101, Taiwan}

\author{Hsien-Ching Chung\,\orcidicon{0000-0001-9364-8858}}
\email[E-mail: ]{hsienching.chung@gmail.com}
\affiliation{RD Dept., Super Double Power Technology Co., Ltd., Changhua City, Changhua County, 500042, Taiwan}

\author{Ming-Fa Lin}
\email[E-mail: ]{mflin@mail.ncku.edu.tw}
\affiliation{Department of Physics, National Cheng Kung University, Tainan 70101, Taiwan}

\date{\today}

\begin{abstract}
The widely use of lithium-ion (Li-ion) batteries in various fields, from portable products to large-scale energy storage systems, has revolutionized our daily life. The 2019 Nobel Prize in Chemistry has been awarded to John B. Goodenough, M. Stanley Whittingham, and Akira Yoshino for their contributions in developing Li-ion batteries. Although Li-ion batteries are currently on-growing research topics, lithium availability is still a problem for mass production. In contrast to lithium, sodium resources are almost unlimited on Earth, and sodium is one of the most abundant elements in the Earth's crust. Hence, sodium-ion (Na-ion) batteries as a counterpart of Li-ion batteries have the potential to serve as the next-generation batteries. In this work, a brief history and recent development of Na-ion batteries are described. The fundamental physical and electronic properties, such as geometric structures, band structure, density of states, and spatial charge distributions, of Na-intercalation compounds are discussed. The outlook of Na-ion batteries is given at the last.
\begin{description}
\item[Usage]
This is a preprint version.
\end{description}
\end{abstract}

\maketitle


\section{Introduction}

The widely use of lithium-ion (Li-ion) batteries in various fields, from 3C products to MWh-class grid-scale energy storage systems (ESSs), has revolutionized our daily life~\cite{BookChChung2021EngIntePotentialAppOutlooksLiIonBatteryIndustry, engrxiv2020OutlooksLiIonBatteriesH.C.Chung}. The 2019 Nobel Prize in Chemistry has been awarded to John B. Goodenough, M. Stanley Whittingham, and Akira Yoshino for their contributions in developing Li-ion batteries. The story can go back to the 1970s, Whittingham demonstrated the prototype framework of Li-ion batteries~\cite{Mater.Res.Bull.10(1975)363M.S.Whittingham, Science192(1976)1126M.S.Whittingham}. However, the commercialization is not smooth. The dendrite of lithium metal leads to short-circuiting and thermal runaway. After a lot of effort, stable full-cell Li-ion batteries are made. In 1985 at Asahi Kasei Corporation, Akira Yoshino~\cite{Pat1989293SecondaryBattery1985A.Yoshino, Angew.Chem.Int.Ed.51(2012)5798A.Yoshino} assembled a full rechargeable battery including the petroleum coke anode with Goodenough's LiCoO$_2$ cathode\cite{Chem.Mater.22(2010)587J.B.Goodenough, Nat.Commun.11(2020)1550A.Manthiram, J.SolidStateChem.71(1987)349A.Manthiram, J.PowerSources26(1989)403A.Manthiram, Chem.Rev.113(2013)6552C.Masquelier, Mater.Res.Bull.18(1983)461M.M.Thackeray, Mater.Res.Bull.15(1980)783K.Mizushima} based on Whittingham's framework~\cite{Mater.Res.Bull.10(1975)363M.S.Whittingham, Science192(1976)1126M.S.Whittingham}, preventing the risk of dendrite-formation-induced thermal runaway and ensuring the stability for the commercial market. This battery was later commercialized by Sony in 1991 with a gravimetric energy capacity of 80 Wh/kg and volumetric energy capacity of 200 Wh/L~\cite{J.PowerSources100(2001)101Y.Nishi}.

Currently, Li-ion batteries have exhibited many branches based on their active materials, performing various applications \cite{BookLinFirstPrinciplesCathodeElectrolyteAnodeBatteryMaterials}. They are commonly named by the cathode (positive) materials, such as lithium cobalt oxide (LiCoO$_2$ or LCO) \cite{J.PowerSources115(2003)171A.D.Pasquier, Mater.Res.Bull.15(1980)783K.Mizushima, J.PowerSources100(2001)101Y.Nishi}, lithium manganese oxide (LiMnO$_2$ or LMO) \cite{Nature381(1996)499A.R.Armstrong, J.SolidStateChem.124(1996)83P.Strobel, Prog.SolidStateChem.25(1997)1M.M.Thackeray, J.Am.Ceram.Soc.82(1999)3347M.M.Thackeray, J.PowerSources21(1987)1M.M.Thackeray}, lithium iron phosphate (LiFePO$_4$ or LFP) \cite{Electrochem.Soc.Interface21(2012)37D.Doughty, J.Electrochem.Soc.144(1997)1188A.K.Padhi, J.Electrochem.Soc.144(1997)1609A.K.Padhi, J.Electrochem.Soc.148(2001)A224A.Yamada}, lithium nickel cobalt manganese oxide (Li(Ni$_x$Co$_y$Mn$_z$)O$_2$ or NCM) \cite{Adv.Mater.18(2006)2330K.M.Shaju, Nat.Mater.8(2009)320Y.K.Sun, J.PowerSources119-121(2003)171N.Yabuuchi}, and lithium nickel cobalt aluminum oxide (LiNi$_{0.8}$Co$_{0.15}$Al$_{0.05}$O$_2$ or NCA) \cite{Energies10(2017)1314G.Berckmans, J.Electrochem.Soc.158(2011)A1115S.K.Martha}. Some are named by their anode (negative) materials, such as lithium titanium oxide (Li$_4$Ti$_5$O$_{12}$ or LTO) \cite{Adv.EnergyMater.5(2015)1402225V.Aravindan, Adv.Funct.Mater.23(2013)959Z.Chen, Adv.Mater.20(2008)2878Y.G.Guo, J.Mater.Chem.A5(2017)6368C.P.Han, J.Electrochem.Soc.151(2004)A2082J.Jiang, Adv.Funct.Mater.23(2013)1214C.Kim, J.PowerSources279(2015)481E.Pohjalainen, Chem.Rev.113(2013)5364M.V.Reddy, Ionics20(2014)601C.P.Sandhya, Chem.Commun.48(2012)516M.S.Song, J.Mater.Chem.A2(2014)631M.S.Song, NewJ.Chem.39(2015)38X.Sun, J.Phys.Chem.Solids71(2010)1236T.F.Yi, J.Mater.Chem.A3(2015)5750T.F.Yi, Mater.Sci.Eng.RRep.98(2015)1B.Zhao, EnergyEnviron.Sci.5(2012)6652G.N.Zhu}. LCO and LMO-based batteries with medium energy capacities and high nominal voltages (3.6$\sim$3.7 Volt) are usually used in portable devices, such as smartphones and laptops. NCM and NCA-based batteries with high energy capacities and high nominal voltages (3.6$\sim$3.7 Volt) are usually used in power applications, such as electric vehicles. LFP-based batteries with low energy capacities, medium nominal voltages (3.2 Volt), and high safety features are usually used in large-scale power and stationary application, such as electric buses and grid-scale energy storage systems. LTO-based batteries with low energy capacities, low nominal voltages (2.4 Volt), and high safety features are usually used in power applications, such as electric vehicles and uninterruptible power supply (UPS) \cite{engrxiv2020OutlooksLiIonBatteriesH.C.Chung, BookChChung2021EngIntePotentialAppOutlooksLiIonBatteryIndustry}.

In 2019, the battery market is USD 25 billion and showing a growing tendency. By 2030 the battery market will be worth USD 116 billion annually according to BloomNEF's forecasts \cite{BatteryMarket156USD2019Bloomberg}, and this doesn't include investment in the supply chain. A decade ago, the Li-ion batteries were a pricey proposition. In 2010, the Li-ion battery packs cost 1183/kWh. Nine years later, the price had decreased nearly tenfold to USD 156/kWh in 2019, falling 87\% in real terms, driving a fast-expanding market for electric vehicles (EVs). Manufacturers are closing in on a point where EVs will approach cost parity with their fossil fuel-powered cousins at around USD 100/kWh. That price is widely seen as a sweet point in the sector, where consumers will no longer regard EVs as pricey options. Cost reductions in 2019 are attributed to increasing order size, growth in EV sales, and the continued penetration of high energy density cathodes. The advance in pack designs and falling manufacturing costs will further drive prices down. BloombergNEF forecast that Li-ion battery costs will fall under 100 USD/kWh in 2024 and hit around 60 USD/kWh by 2030 \cite{BatteryMarket156USD2019Bloomberg}.

Although Li-ion batteries currently give an appropriate solution to overcome the challenges in realizing sustainable energy development, reflecting on the growing EV market and emerging ESS market, the lithium reserves must be taken into consideration \cite{Nat.Chem.2(2010)510J.M.Tarascon}. According to the U.S. Geological Survey (USGS), lithium has historically been acquired from either continental brines or hard-rock minerals. Chile has been a leading producer of lithium carbonate (Li$_2$CO$_3$) for a long time, with production from two Salar de Atacama (Atacama Salt Flat) brine operations next to the Andes Mountains in South America \cite{Resour.Conserv.Recycl.174(2021)105762J.C.Kelly}. Although lithium markets vary by location, global end-use markets are estimated as follows: batteries are the largest (74\%), ceramics and glass (14\%) and the rest are lubricating greases (3\%), continuous casting mold flux powders (2\%), polymer production (2\%), air treatment (1\%), and other uses (4\%). Lithium consumption for batteries has largely grown in recent years owing to rechargeable Li-ion batteries used extensively in the raising market for EVs, portable electronic devices, and grid-scale storage applications \cite{USGSMineralCommoditySummaries2022}. Excluding U.S. production, worldwide lithium production in 2021 increased by 21\% to approximately 100,000 tons from 82,500 tons in 2020 reflects strong demand from the Li-ion battery market and increased prices of lithium. Global consumption of lithium in 2021 was estimated to be 93,000 tons, a 33\% increase from 70,000 tons in 2020. Lithium resources are unevenly distributed, the mine production of the first two countries are Australia (55,000 tons) and Chile (26,000 tons) \cite{USGSMineralCommoditySummaries2022}. In contrast to lithium, sodium resources are almost unlimited on Earth, and sodium is one of the most abundant elements in the Earth's crust. The sodium resources can be easily found in the ocean. Sodium can easily be obtained by evaporation of seawater (11,000 mg/L in seawater) where the lithium content in seawater is much lower than that of sodium (0.18 mg/L) \cite{BeilsteinJ.Nanotechnol.6(2015)1016P.Adelhelm}. Additionally, sodium is the second-lightest and second-smallest alkali metal below lithium in the periodic table. Based on material abundance and standard electrode potential, rechargeable sodium-ion batteries (or Na-ion batteries) are the potential alternative to Li-ion batteries \cite{Adv.EnergyMater.2(2012)710S.W.Kim, Adv.EnergyMater.21(2011)3859S.Komaba, Adv.Funct.Mater.23(2013)947M.D.Slater}.

Na-ion batteries are operable at ambient temperature, and metallic sodium is not used as the anode (negative) electrode, which is different from other commercialized high-temperature sodium-based technology, such as sodium-sulfur batteries (Na/S batteries) \cite{Int.J.Appl.Ceram.Technol.1(2004)269T.Oshima} and Na/NiCl$_2$ \cite{J.Electrochem.Soc.136(1989)1274R.J.Bones} batteries. These batteries apply alumina-based solid (ceramic) electrolyte under high operation temperature (250$\sim$350$^\circ$C) for maintaining the electrodes in the liquid state to ensure good contact with the solid-electrolyte. Because molten sodium and sulfur are adopted as active materials at such high temperatures, safety issues remain a critical problem for consumer appliances. On the contrary, Na-ion batteries composed of sodium insertion materials with polar aprotic solvent as an electrolyte are free from metallic sodium and the safety issues of high temperature. Structures, components, systems, and charge storage mechanisms of Na-ion batteries are essentially analogous to those of Li-ion batteries \cite{Adv.EnergyMater.21(2011)3859S.Komaba}. Na-ion batteries are composed of two sodium insertion materials, cathode (positive) and anode (negative) electrodes, which are electronically separated by electrolyte as a pure ionic conductor. The battery performance relies on selected battery components, and many different Na-ion batteries for different purposes can be fabricated.

The abundance of materials is a straightforward reason for considering sodium ions as the charge carriers in secondary rechargeable batteries. There is an obvious disadvantage when compared between Li and Na metal electrodes, i.e., the theoretical gravimetric capacities of Na (1166 mAh/g) are much less than that of Li (3861 mAh/g), also the volumetric capacities of Na (1131 mAh/g) are much less than that of Li (2062 mAh/g) \cite{Chem.Rev.114(2014)11636N.Yabuuchi}. It seems that if treated Na$^+$/Na as an electrochemical equivalent of Li$^+$/Li, Na is more than three times heavier than Li. However, when compared to the layered cobalt oxides of Na and Li (i.e., NaCoO$_2$ and LiCoO$_2$), the difference between their theoretical capacity becomes smaller. The theoretical capacity is 235 and 274 mAh/g respectively for NaCoO$_2$ and LiCoO$_2$, as one-electron redox of the cobalt ion is assumed to happen (Co$^{3+}$/Co$^{4+}$ redox). In this case, the capacity is decreased by an acceptable 14\%. Another sacrifice can be found in the difference in working voltage range. The working voltage range is 2$\sim$3.5 V \cite{SolidStateIonics3-4(1981)165C.Delmas} for NaCoO$_2$, which is much lower than the 3$\sim$4.2 V for LiCoO$_2$ \cite{Mater.Res.Bull.15(1980)783K.Mizushima}. If the final goal is to establish the energy storage technology based on sodium ions instead of sodium metal, the sacrifice in energy density can be potentially eliminated. Hence, Na-ion batteries are expected to be the alternative battery system for Li-ion batteries.

Early-stage studies of Li$^+$ and Na$^+$ ions as charge carriers for electrochemical energy storage at ambient temperature begun in 1970s. In 1980, lithium cobalt oxide (LiCoO$_2$), which is a lithium-containing layered structure, demonstrated the electrode performance for high-energy positive electrode materials in Li-ion batteries \cite{Mater.Res.Bull.15(1980)783K.Mizushima}. On the other side, sodium cobalt oxide (Na$_x$CoO$_2$) as sodium-containing layered oxides were also reported \cite{SolidStateIonics3-4(1981)165C.Delmas}. The early history of sodium intercalation materials was reviewed and published in 1982 \cite{SolidStateIonics7(1982)199K.M.Abraham, Rev.Chim.Miner.19(1982)343C.Delmas}. However, in the past three decades, important research efforts have been conducted only for Li-ion batteries, and studies on the counterpart sodium intercalation materials for energy storage once almost disappeared.

Two reasons stand for the disappear studies. The first reason is the available energy density. Li-ion batteries were believed to possess higher energy densities than their Na-ion counterpart. Although both systems have similar crystal lattice construction by sheets of edge-sharing CoO$_6$ octahedra, the voltage of NaCoO$_2$ at the end of discharge is lower than that of LiCoO$_2$. When both systems are charged to $>100$ mAh/g, the voltage difference is dropped to approximately 0.4 V \cite{SolidStateIonics3-4(1981)165C.Delmas, Mater.Res.Bull.15(1980)783K.Mizushima}, similar to the difference in the standard electrochemical potential of Li (3.040 V) \cite{Nat.Nanotechnol.12(2017)194D.Lin, Joule2(2018)833B.Liu, EnergyEnviron.Sci.7(2014)513W.Xu} and Na (2.71 V). The voltage difference becomes more critical as Na is a major component in the structure. Consequently, the available energy density of the Na system is much lower than that of the Li system. Especially, when the same chemistry is used, such as redox species and host crystal structures. The comparison between NaCoO$_2$ and LiCoO$_2$ is the typical case.

The second reason was the lack of appropriate anode (negative) electrodes for a long period of time. From the 1980s, carbonaceous materials have been found to be potential candidates for Li intercalation hosts. Carbonaceous materials are currently widely adopted as anode (negative) electrode materials for Li-ion batteries, such as carbon fiber \cite{J.PowerSources26(1989)535R.Kanno}, pyrolytic carbon \cite{J.PowerSources26(1989)545M.Mohri}, and graphite \cite{J.Electrochem.Soc.137(1990)2009R.Fong, J.Electrochem.Soc.140(1993)2490T.Ohzuku}. The research interest in Li-ion batteries is further accelerated by the application of graphite, resulting in high capacity (theoretically 372 mAh/g). Unfortunately, graphite is not suitable for an intercalation host of sodium ions, e.g., Ge and Fouletier (35 mAh/g) \cite{SolidStateIonics28(1988)1172P.Ge}, Doeff et al. (93 mAh/g) \cite{Electrochem.Solid-StateLett.6(2003)A1J.Barker, J.Electrochem.Soc.140(1993)L169M.M.Doeff}, Thomas et al. (55 mAh/g) \cite{Electrochim.Acta45(1999)423P.Thomas}. Before the 1990s, several studies were available as potential anode (negative) electrode materials for Na-ion batteries, e.g., Na-Pb alloy \cite{J.Electrochem.Soc.134(1987)1730T.R.Jow, J.Electrochem.Soc.140(1993)2726Y.P.Ma} and disordered carbon \cite{J.Electrochem.Soc.140(1993)L169M.M.Doeff}. However, it was obvious that the energy density of Na-ion batteries with these anode materials was inferior to that of Li-ion batteries with graphite anodes.

As the first turning point in the Na-ion battery research field, in 2000, Stevens and Dahn reported a high capacity of 300 mAh/g for Na-ion batteries with hard carbon \cite{J.Electrochem.Soc.147(2000)1271D.A.Stevens}, close to that for Li-ion batteries. Hard carbon is now extensively studied as a potential candidate for anode (negative) electrode material for Na-ion batteries. As the second important finding, in 2006, Okada et al. reported that NaFeO$_2$ is electrochemically active in Na-ion batteries based on the Fe$^{3+}$/Fe$^{4+}$ redox couple \cite{210thECSMeetingAbstracts2006S.Okada}. The capacity of NaFeO$_2$ is 80 mAh/g, and the 3.3 V flat discharge profile is similar to that of isostructural LiCoO$_2$ in Li-ion batteries. The Fe$^{3+}$/Fe$^{4+}$ redox is unique chemistry for the Na-ion batteries and never demonstrated as active for LiFeO$_2$ in the Li-ion batteries (since layered rocksalt LiFeO$_2$ is metastable in Li-ion batteries).

\section{Recent development}

The Na-ion batteries with organic electrolytes are state-of-the-art technology. The development of renewable and environment-friendly organic-electrolyte-based Na-ion batteries has attracted researchers' attention in the past decade \cite{BeilsteinJ.Nanotechnol.6(2015)1016P.Adelhelm, Adv.EnergyMater.8(2018)1702869A.Bauer}. In organic-electrolyte-based Na-ion batteries, the electrolytes are synthesized by dissolving one of the ionic sodium salts (e.g., NaClO$_4$, NaPF$_6$, and NaTFSI) in non-aqueous solvents (e.g., EC, PC, DMC, DEC, DME), like in Li-ion batteries \cite{Int.J.HydrogenEnergy41(2016)2829K.Vignarooban}. The ionic conductivity of an electrolyte synthesized with 1 M NaClO$_4$ in EC:DME (50:50 wt\%) approaches to 12.5 ms/cm (almost the same value as 1 M LiPF$_6$ in EC:DMC (50:50 wt\%)) \cite{Int.J.HydrogenEnergy41(2016)2829K.Vignarooban}.

The energy and power density of the Na-ion batteries is directly related to the cathode materials. Analogous to Li-ion batteries, sodium cathodes can be divided into three major groups: layered, olivine (NASICON), and spinel. Currently, layered structure oxides \cite{Adv.EnergyMater.8(2018)1701610J.Deng, ChemSusChem7(2014)2115S.Guo, J.Mater.Chem.A6(2018)8558A.Konarov, Chem.Commun.51(2015)4693H.Liu, EnergyEnviron.Sci.10(2017)1051N.Ortiz-Vitoriano} and NASICON structure phosphates have demonstrated potential performance \cite{Nat.Commun.10(2019)1480M.Chen}. Recently, Prussian blue analogs (PBAs) (Na$_2$M[Fe(CN)$_6$], where M = Co, Mn, Ni, Cu, Fe, etc.) have been extensively studied based on their large vacancies in lattice space structure, providing many sites and transport channels for reversible Na-ion deintercalation \cite{iScience3(2018)110B.Wang, Adv.EnergyMater.8(2018)1701785Y.You}.

Layered oxides are possible candidates as cathode materials for Na-ion batteries, possessing a common formula Na$_x$XO$_2$, where X is one or several transition metals (Mn, Fe, Co, Ni, Ti, V, Cr, etc). 2D layered sodium transition metal oxides are extensively studied based on their electrochemical features. The 2D layered sodium transition metal oxides are classified as an O3 type and P2 type structure, where the O and P indicate the location of Na ion in the crystal structure (i.e., P-prismatic site and O-octahedral site), and the numbers indicate the transition metal layer in the repeating unit cell in the structure \cite{Adv.EnergyMater.8(2018)1703137C.Delmas}. O3- and P2‐types are the common structural polymorphs of layered transition metal oxides.

In the P2-type materials, manganese-based (Na$_{2/3}$MnO$_2$) cathode has attracted much attention based on the low cost of manganese, and it gives high discharge capacity ($>150$ mAh/g) \cite{J.Nanopart.Res.20(2018)160X.Zhu}. However, the manganese leads to structural distortions as the Mn$^{3+}$ are dominant in the structure. This phenomenon can be attributed to Jahn-Teller distortion, causing elongation or compression in the $z$-axis. The anisotropic changes in the lattice parameters during charge and discharge result in fast capacity fading \cite{Chem.Commun.51(2015)4693H.Liu}. One of the possible routes to overcome the problem is to dilute the Mn$^{3+}$ concentration in the crystal structure and decrease the anisotropic change by substituting it with other metal cations \cite{Funct.Mater.Lett.11(2018)1830003I.ElMoctar, Adv.Mater.29(2017)1701968J.Z.Guo, J.Electrochem.Soc.165(2018)A4058E.Irisarri, Chem.Mater.27(2015)7258H.Kim}.

Clément et al. \cite{EnergyEnviron.Sci.9(2016)3240R.Clement} studied the effect of Mg doping on the P2-type Na$_{2/3}$MnO$_2$ compound by varying the amount of dopant, i.e., Na$_{2/3}$Mn$_{1-y}$Mg$_y$O$_2$ ($y = 0$, 0.05, and 0.1). They demonstrated that Mg substitution results in smoother electrochemistry, with fewer distinct electrochemical processes, improving rate performance, and better capacity retention. Mg doping reduces the number of Mn$^{3+}$ Jahn-Teller centers and delays the high voltage phase transition occurring in P2-Na$_{2/3}$MnO$_2$. The 5\% Mg-doped phase exhibited the highest capacity and rate performance. Kang et al. \cite{J.Mater.Chem.3(2015)22846W.Kang} demonstrated copper-substituted P2-type Na$_{0.67}$Cu$_x$Mn$_{1-x}$O$_2$ ($x = 0$, 0.14, 0.25, and 0.33) compound to enhance the rate performance of P2-type Na$_{0.7}$MnO$_2$. The materials show excellent stability, retaining more than 70\% of the initial capacity after 500 cycles at 1000 mA/g.

In NASICON type of cathodes, Na$_3$V$_2$(PO$_4$)$_3$ and its derivatives demonstrated great electrochemical performance \cite{Adv.Mater.29(2017)1701968J.Z.Guo}. Although the capacity is less than that of the layered oxides, the operating potential makes it more promising. However, the cost and toxicity of the vanadium restrict its next step and widespread to the market. Many research works \cite{Front.Chem.8(2020)152M.Chen, J.Electrochem.Soc.165(2018)A4058E.Irisarri, Adv.EnergyMater.8(2018)1701785Y.You} have reported systematic studies of the electrochemical performance of various cathode materials. Considerable comparison can be obtained between the cycling profiles for these materials, realizing the cathode behavior.

\section{Fundamental physical and electronic properties of Na-intercalation compounds}

The fundamental physical and electronic properties, such as geometric structure, band structure, density of states, and spatial charge distribution, of Na-intercalation compounds (NaC$_8$, NaC$_{18}$, NaC$_{24}$, and NaC$_{32}$) are discussed.

\subsection{Geometric structure}

Generally, it is well-known that graphite intercalation compounds exhibit the planar carbon-honeycomb symmetries and the well-characterized intercalant distributions, especially for metal atoms or even for large molecule ones \cite{Phys.Chem.Chem.Phys.19(2017)7980P.Bhauriyal}. The graphitic layers are attracted by the weak but significant van der Waals interactions, leading to the easy modulations by the adatom \cite{Phys.Rev.B84(2011)241404C.A.Howard} or large-molecule intercalations/de-intercalations \cite{Phys.Rev.Appl.12(2019)044060Q.P.Wang}. This clearly indicates strong $\sigma$ bondings of graphitic layers, as well as the non-covalent interlayer and intra-intercalant interactions. From the delicate analyses of VASP simulations, Na-intercalations could be classified into stable and quasi-stable configurations according to their ground state energies, $E_{gs}$.

\begin{figure*}
  \includegraphics[]{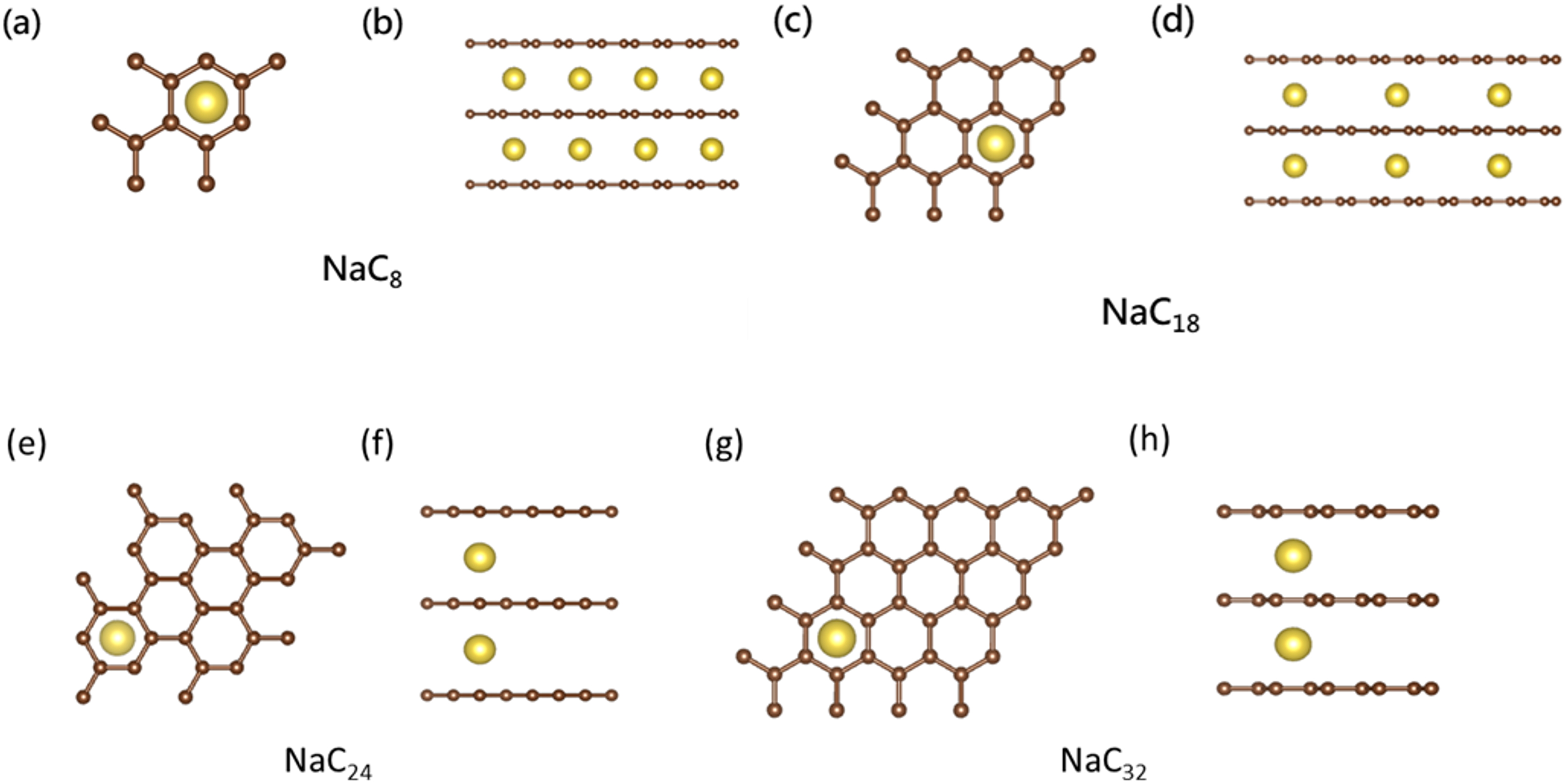}\\
  \caption{
(Color online)
The optimal crystal structures of sodium-graphite intercalation compounds with the top/side view (the left-/right-hand sides): (a)/(b) NaC$_8$, (c)/(d) NaC$_{18}$, (e)/(f) NaC$_{24}$, (g)/(h) NaC$_{32}$.
}
  \label{fig:Figure01}
\end{figure*}

For the stage-1 NaC$_x$, where $x= 8$, 18, 24, and 32 (as shown in Fig.~\ref{fig:Figure01}), the hollow-site position possesses the lowest ground state energy, that is to say, the hollow-site position is the most stable geometric structure. The interlayer distances of NaC$_8$, NaC$_{18}$, NaC$_{24}$, and NaC$_{32}$ are, respectively, 4.600, 4.511, 4.452, and 4.411 {\AA}; also, the Na-C bond lengths are, respectively, 2.716, 2.668, 2.651, and 2.627 {\AA}. The calculated results reveal that the interlay distances and Na-C bond lengths decrease with the lower concentrations. However, the C-C bond lengths are influenced very slightly (details shown in Table~\ref{tab:table01}).

\begin{table*}
\caption{
\label{tab:table01}
The calculated results of NaC$_8$, NaC$_{18}$, NaC$_{24}$, and NaC$_{32}$.
}
\begin{ruledtabular}
\begin{tabular}{ccccccc}
 & Ground state energy (eV) & intralayer C-C ({\AA}) & Na-C ({\AA}) & Na-Na ({\AA}) & Interlayer distance ({\AA}) & Blue shift of $E_F$ (eV)\\
\hline
NaC$_8$ & -75.017 & 1.434 & 2.716 & 4.985 & 4.600 & 1.48\\
NaC$_{18}$ & -169.527 & 1.129 & 2.688 & 7.431 & 4.511 & 1.17\\
NaC$_{24}$ & -225.407 & 1.427 & 2.651 & 8.572 & 4.452 & 1.02\\
NaC$_{32}$ & -299.823 & 1.427 & 2.627 & 9.891 & 4.411 & 0.92
\end{tabular}
\end{ruledtabular}
\end{table*}

The rich crystal symmetries of 3D graphite intercalation compounds, which are clearly revealed in the graphitic and intercalant layers, as well as, their observable spacing, could be verified by the high-resolution measurements of X-ray diffraction peaks \cite{EnergyEnviron.Sci.12(2019)3575H.J.Liang, PowderDiffr.32(2017)S43J.C.Pramudita, Electrochim.Acta54(2009)5648T.E.Sutto}, reflection electron diffraction patterns (top views) \cite{Appl.Phys.Lett.52(1988)103H.E.Elsayed-Ali, e-J.Surf.Sci.Nanotechnol.16(2018)88Y.Horio} and tunneling electronic microscopic spectra (side views) \cite{Phys.Rev.B97(2018)125401W.Ko}, but not those by scanning tunneling microscopy (STM) \cite{Phys.Rev.B50(1994)1839Z.Y.Rong, Phys.Rev.B48(1993)17427Z.Y.Rong}. The former is frequently utilized in these layered materials, e.g., strong evidence of stage-n alkali graphite intercalation compounds Li$_{6n}$ and MC$_{8n}$, where M = Na, K, Rb, and Cs.

\subsection{Band structure}

Pristine graphene, a single carbon-honeycomb crystal, is a well-known semiconductor with zero density of states at the Fermi level \cite{Nature430(2004)870A.Hashimoto}. A pair of linear and isotropic valence and conduction subbands intersect there so that a well-defined Dirac-cone structure is initiated from the K/K' valley. This feature is modified by the perpendicular stacking configurations through the weak, but important van der Waals interactions. The linear and/or parabolic energy dispersions come to exist at the K and H valleys, depending on the stacking symmetries. Most importantly, the significant band overlaps, corresponding to the semi-metallic behaviors, are purely induced by the interlayer C-2$p_z$-orbital hybridizations. The AA-/ABC-stacked graphite has the highest/lowest densities of free conduction electrons/valence holes. The semiconductor-semimetal transition would become semimetal-metal after the $n$- or $p$-type dopings.

\begin{figure*}
  \includegraphics[]{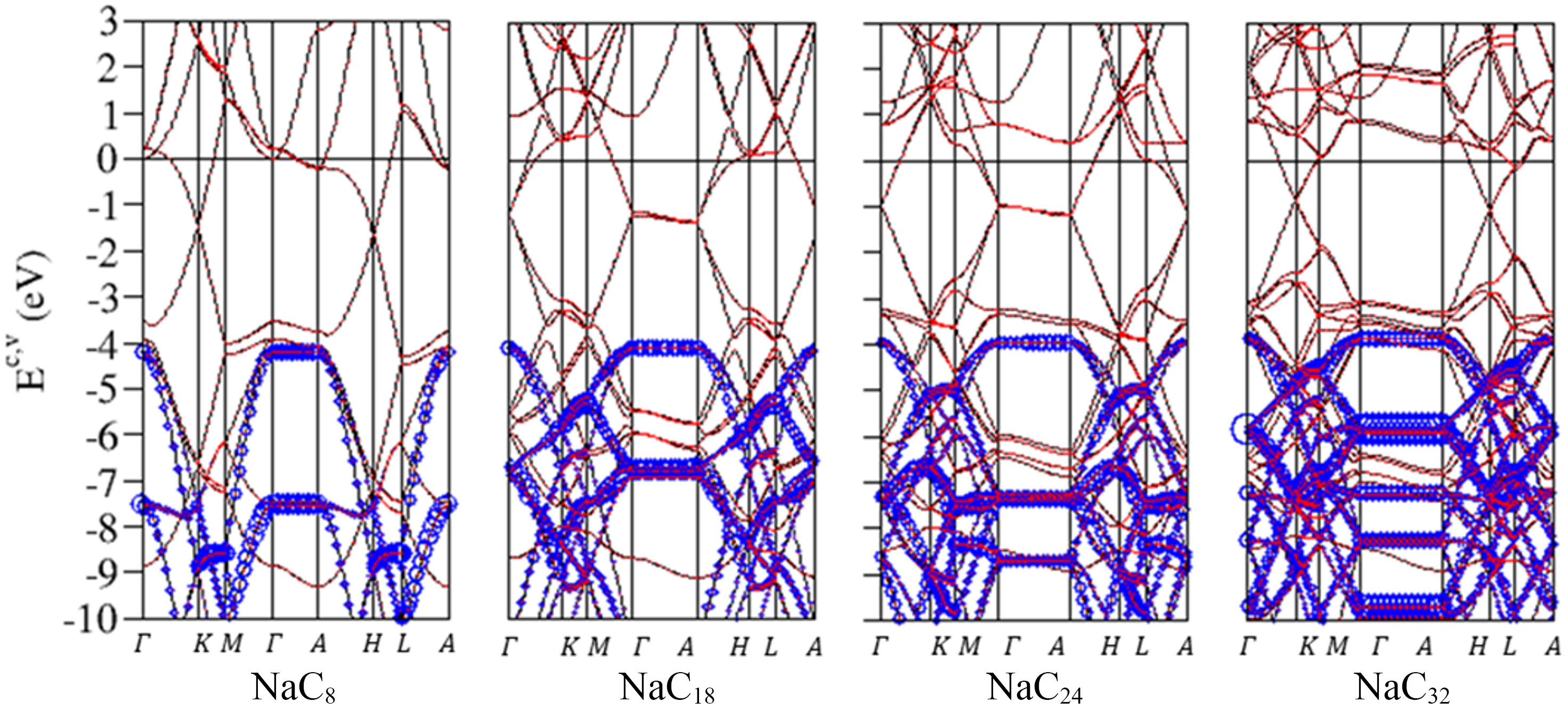}\\
  \caption{
(Color online)
Band structures of Na-based graphite intercalation compounds within the active valence energy spectra, with atom dominances (red/blue for Na/C): (a) NaC$_8$, (b) NaC$_{18}$, (c) NaC$_{24}$, and (d) NaC$_{32}$.
}
  \label{fig:Figure02}
\end{figure*}

Band structures of sodium-graphite intercalation compounds, as clearly displayed in Fig.~\ref{fig:Figure02} within a sufficiently wide energy range of $-10~\mathrm{eV} \leq E^{c,v} \leq 3~\mathrm{eV}$, respectively, are fully supported by the carbon- and intercalant-atom dominances. Their main features cover the number of valence subbands (the active atoms and orbitals in a unit cell) \cite{BookLin2020SiliceneBasedLayeredMaterials, BookLin2017.Geo.Elec.Prop.Graphene-RelatedSys.}, the enhanced asymmetry of valence and hole conduction electron energy spectra about the Fermi level (the intercalant-carbon interlayer interactions) \cite{Phys.Lett.A352(2006)446J.H.Ho}, the blue shift of $E_F$ (the semimetal-metal transitions due to the charge transfer of metal adatoms) \cite{Phys.Rev.B83(2011)121201Y.H.Ho}, the dependence of $E_F$ on the chemical intercalations (the densities of free conduction electrons under the various intercalation cases), the slight/major modifications of carbon 2$p_z$-$\pi$ and [2$s$, 2$p_x$, 2$p_y$]-$\sigma$ bands along the different wave-vector paths, the subband non-crossing/crossing/anti-crossing behaviors \cite{BookLin2017.Opt.Prop.GrapheneMag.Elec.Fields}, the diverse energy dispersions near the high-symmetry points \cite{BookLin2019.DiverseQuanti.PhenomenaLayeredMater.}, and their band-edge states (the critical points in energy-wave-vector spaces with zero group velocities) \cite{BookLin2018StructureAdatomEnrichedPropertiesGNRs}. The Fermi level is mainly determined by the concentration, distribution configuration, and charge transfer of intercalant metal atoms. Its characteristics of the Fermi-Dirac distribution at low temperatures strongly affect the other essential properties, especially for single- and many-particle optical absorption spectra and Coulomb excitations. According to the band structures, the fermi level intersects the conduction bands; that is, the sodium-graphite intercalation compounds exhibit features of $n$-type doping semiconductors.

\subsection{Density of states}

In the density of states (DOS), the critical points presented in the energy-wave-vector space are responded to the unusual van Hove singularities. Their special structures are diversified by the different band-edge states of the parabolic, linear, and partially flat energy dispersions. The calculated results are delicately decomposed into the specific contributions of active atoms and orbitals. As a result, significant orbital hybridizations are achieved from the merged unusual structures at the specific energies. These detailed analyses could be generalized to other complicated condensed-matter systems. The current predictions mainly depend on the intercalant concentrations, e.g., the different blue shifts of the Fermi level in sodium-graphite intercalation compounds.

\begin{figure*}
  \includegraphics[]{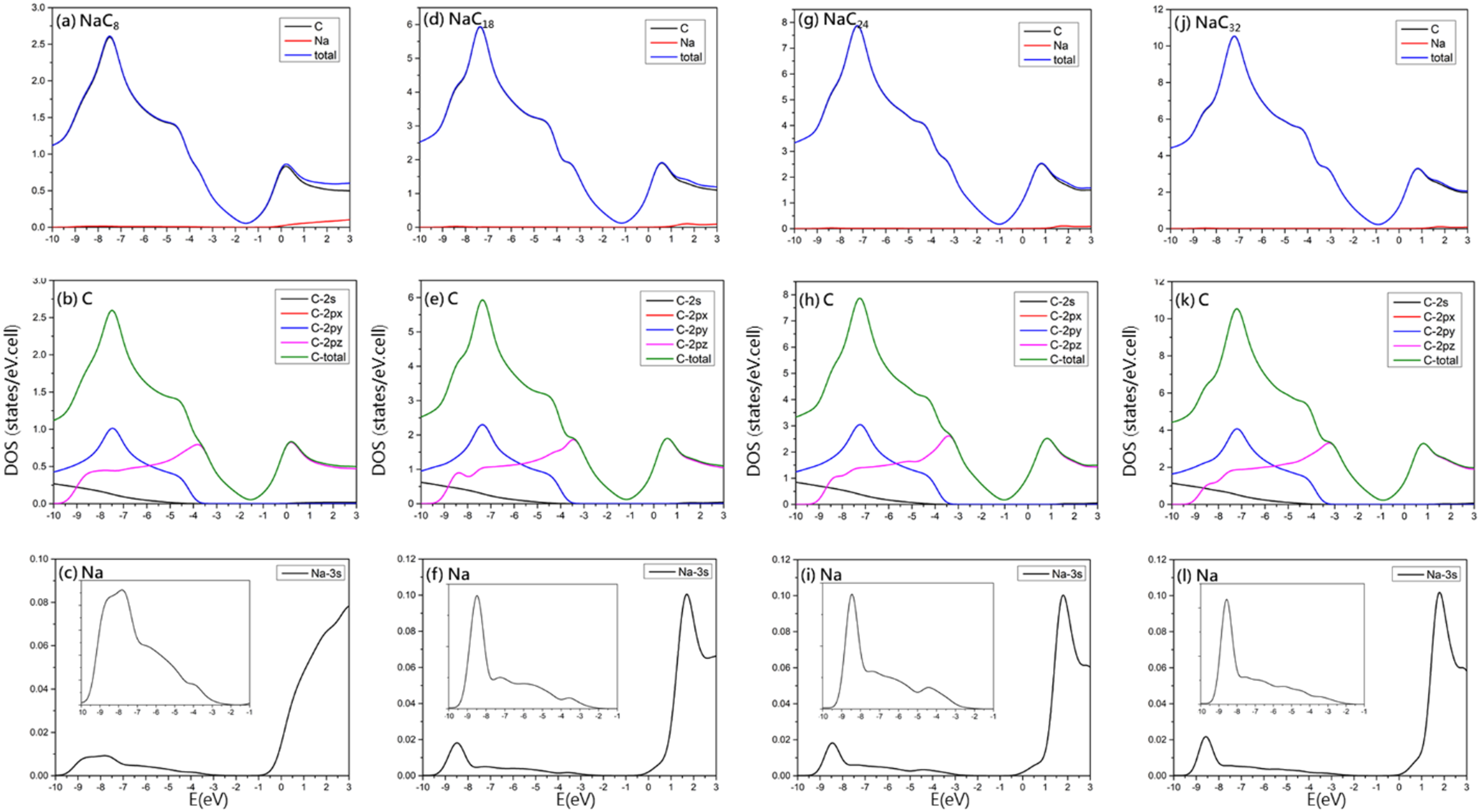}\\
  \caption{
(Color online)
The density of states in Na-graphite intercalation compounds: (a)/(b)/(c) NaC$_8$, (d)/(e)/(f) NaC$_{18}$, (g)/(h)/(i) NaC$_{24}$, (j)/(k)/(l) NaC$_{32}$.
}
  \label{fig:Figure03}
\end{figure*}

The atom- and orbital-decomposed van Hove singularities are very suitable in determining the active orbital hybridizations of carbon-carbon, intercalant-intercalant, and carbon-intercalant bonds. In general, a Moire superlattice has a lot of carbon atoms with four significant orbitals, compared with a single metal atom with one effective orbital sodium. This leads to the dominating/minor contributions from carbon-honeycomb lattices/the intercalant layers under any intercalation cases (as shown in Fig.~\ref{fig:Figure03}). The main features of van Hove singularities mainly arise from the modified $\pi$- and $\sigma$-electronic energy subbands. A local minimum density of states, a dip structure, corresponds to the corrected Dirac-cone structure, and its energy difference with the Fermi level of $E_F=0$ is an observable blue shift associated with the electron transfer of metal adatoms. The important values (i.e., the blue shift of $E_F$) of NaC$_8$, NaC$_{18}$, NaC$_{24}$, and NaC$_{32}$ are estimated to be, respectively, 1.48, 1.17, 1.02, and 0.92 eV. It displays that the higher concentrations of Na will provide more electrons transferring from sodium atoms to carbon atoms. Their magnitudes, which correspond to the 3D free electron densities, will have strong effects on the carrier excitations and transports \cite{BookLin2017.TheoryMagnetoelectricProp.2DSys., BookLin2019.DiverseQuanti.PhenomenaLayeredMater.}, e.g., the optical plasmon modes of conduction electrons in the $n$-type graphite intercalation compounds \cite{BookLin2017.TheoryMagnetoelectricProp.2DSys., BookLin2019.DiverseQuanti.PhenomenaLayeredMater.}.

According to four-orbital dominances of carbon atoms, the main features in the density of states could be classified into four specific energy ranges: e.g., for NaC$_8$: Firstly, the $\sigma$-electronic energy spectrum of C-2$s$ orbitals initiated from $E \sim -5.0$ eV; secondly, the $\sigma$-valence subbands due to C-[2$p_x$, 2$p_y$] orbitals driven from $E \sim -3.5$ eV and dominant within $-10.0 \mathrm{eV} \leq E \leq -6.0 \mathrm{eV}$; thirdly, the $\pi$-electronic valence spectrum of C-2$p_z$ orbitals in $-10.0 \mathrm{eV} \leq E \leq -6.0 \mathrm{eV}$ with dominance above $-6.0$ eV; fourthly, the $\pi^*$ conduction spectrum higher than $-1.50$ eV. There exist three obvious van Hove singularities, corresponding to the dominant carbon-2$p_z$-orbital contributions, respectively, at $-3.5$ eV, $-1.5$ eV, and $0.2$ eV. Furthermore, the sodium 2$s$-orbital contributions are merged with them, clearly illustrating the 2$p_z$-2$s$ hybridization in each Na-C band. Similar chemical bondings are revealed in NaC$_{18}$, NaC$_{24}$, and NaC$_{32}$. The intercalant-decomposed contributions show the reduced amplitudes of van Hove singularities as their concentration decreases. This indicates 3$s$-3$s$ single-orbital hybridizations for Na-Na bonds. The featured van Hove singularities have successfully identified the active orbital hybridizations of chemical bonds, being consistent with the previous development of the quasiparticle framework \cite{RSCAdv.10(2020)23573W.B.Li}.

\subsection{Spatial charge distribution}

\begin{figure*}
  \includegraphics[]{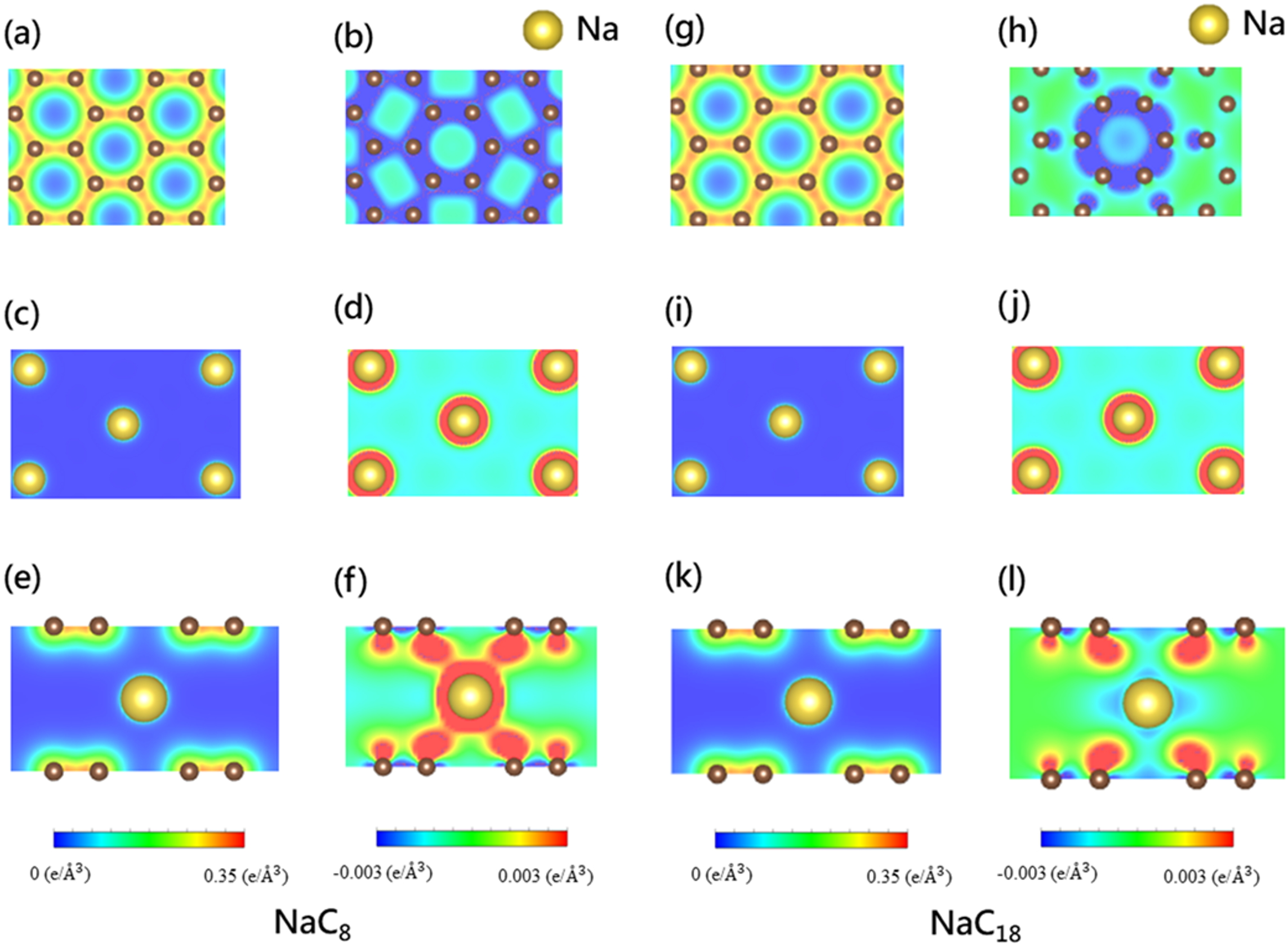}\\
  \caption{
(Color online)
The spatial charge density distributions $\rho (\mathbf{r})$/the variations before and after the chemical intercalations $\Delta \rho (\mathbf{r})$, with the top/side views. The top view on the graphene plane: (a)/(b) NaC$_8$, (g)/(h) NaC$_{18}$. The top view on the Na plane: (c)/(d) NaC$_8$, (i)/(j) NaC$_{18}$. The side view: (e)/(f) NaC$_8$, (k)/(l) NaC$_{18}$.
}
  \label{fig:Figure04}
\end{figure*}

The spatial charge density ($\rho (\mathbf{r})$) and its variation ($\Delta \rho (\mathbf{r})$) after intercalation (as shown in Fig.~\ref{fig:Figure04}) are clearly illustrated in a unit cell through the top- and side-views under the different chemical environments. The calculated results could be further utilized to test the X-ray diffraction peak structures \cite{J.Mater.Eng.Perform.7(1998)329S.J.Kerber, Acc.Chem.Res.50(2017)2737J.Li}. By the delicate observations, the critical chemical bondings could be examined from the carbon-honeycomb lattices, the intercalant layers, and their significant spacings. The multi-/single-orbital hybridizations in different chemical bonds are determined by unifying the relevant quantities; that is, the critical quasiparticle pictures are reached through the full cooperation of the atom-dominated band structures, the charge density distributions, and the orbital-decomposed van Hove singularities. These will be very useful in establishing the concise phenomenological models, e.g., the tight-binding model \cite{Phys.Rev.B82(2010)245412S.Konschuh, BookLin2019CoulombExcitationsDecaysGrapheneRelatedSystems, Phys.Rev.B39(1989)12520W.Matthew}/the generalized tight-binding model \cite{BookLin2017.TheoryMagnetoelectricProp.2DSys., Int.J.QuantumChem.95(2003)394X.G.Zhang} in the absence/presence of a perpendicular magnetic field for the rich magnetic quantization phenomena \cite{BookLin2017.TheoryMagnetoelectricProp.2DSys., BookLin2019.DiverseQuanti.PhenomenaLayeredMater., BookLin2019CoulombExcitationsDecaysGrapheneRelatedSystems}.

From carrier distributions about each carbon-honeycomb lattice, the very strong $\sigma$ bondings and the significant $\pi$ ones, are respectively, observed in $(x, y)$ and $(x, z)/(y, z)$ planes. For example, $\rho (\mathbf{r})$ and $\Delta \rho (\mathbf{r})$ of NaC$_8$ present the high charge density between neighboring carbon atoms (the red color between neighboring carbons) and the peanut-like profile (the light yellow-green color). The former is hardly affected by the sodium-atom intercalations. However, the latter is modified by the interlayer Na-C interactions, as clearly illustrated by the obvious changes of carrier density after interactions (the red color near atoms in Fig.~\ref{fig:Figure04}). The detailed analyses further show the specific orbital hybridization of 3$s$-2$p_z$ orbitals in Na-C bonds. Moreover, 3$s$-3$s$ mixings in Na-Na bonds are also examined from the intercalant layers, e.g., the significant charge density deformation of 3$s$-orbital at the height of $z=3.50$ {\AA}. Apparently, similar C-C, C-intercalant, and intercalant-intercalant bondings are revealed in the different concentration cases.

On the experimental examinations, the spatial charge density distributions could be directly deduced from the close co-operation of measured X-ray patterns \cite{Phys.Rev.Lett.79(1997)1285H.Baltes, J.Mater.Eng.Perform.7(1998)329S.J.Kerber, Acc.Chem.Res.50(2017)2737J.Li} and elastic scattering predictions associated with atoms and active orbitals in a unit cell \cite{J.Mater.Eng.Perform.7(1998)329S.J.Kerber, Acc.Chem.Res.50(2017)2737J.Li}. For example, a sodium chloride crystal \cite{J.Phys.Chem.Lett.11(2020)3821K.A.Tikhomirova}, with very strong ionic bonds, is thoroughly investigated and presents an almost isotropic charge distribution near both Na$^+$-cations and Cl$^-$-anions. Similar analyses are available in exploring those of metal-atom graphite intercalation compounds with the concise orbital hybridizations, e.g., the X-ray tests on the theoretical predictions in Fig.~\ref{fig:Figure04} for Na-graphite intercalation compounds. These investigations will determine the active orbital hybridizations from the intralayer \cite{RSCAdv.10(2020)23573W.B.Li, arXiv2001.02042W.B.Li} and interlayer \cite{RSCAdv.10(2020)23573W.B.Li, arXiv2001.02042W.B.Li} atomic interactions. Such works are very difficult to be done for large-molecule cases. Since the X-ray diffraction spectrum only depends on the whole distribution of charge density, the VASP results should be useful in determining it. How to link them is an open calculation issue.

\section{Outlook of Na-ion batteries}

Several aspects might take account for the future development of Na-ion batteries, such as low dimensional materials as potential additives, as well as reuse and recycling of repurposing batteries. The discovery of graphene \cite{Science306(2004)666K.S.Novoselov, Proc.Natl.Acad.Sci.U.S.A.102(2005)10451K.S.Novoselov} and its incredible essential properties (such as high carrier mobility at room temperature ($>200000$ cm$^2$/Vs) \cite{Science312(2006)1191C.Berger, SolidStateCommun.146(2008)351K.I.Bolotin, Phys.Rev.Lett.100(2008)016602S.V.Morozov}, superior thermoconductivity (3000$\sim$5000 W/mK) \cite{NanoLett.8(2008)902A.A.Balandin, Phys.Rev.Lett.100(2008)016602S.V.Morozov}, extremely high modulus (1 TPa) \cite{NanoLett.8(2008)2458J.S.Bunch, Nat.Nanotechnol.6(2011)543S.P.Koenig, Science321(2008)385C.Lee} and tensile strength (130 GPa) \cite{Science321(2008)385C.Lee, Prog.Mater.Sci.90(2017)75D.G.Papageorgiou}, high transparency to incident light over a broad range of wavelength (97.7\%) \cite{Nat.Nanotechnol.5(2010)574S.Bae, Science320(2008)1308R.R.Nair}, anomalous quantum Hall effect \cite{Nature438(2005)197K.S.Novoselov, Science315(2007)1379K.S.Novoselov, Nat.Phys.2(2006)177K.S.Novoselov, Nature438(2005)201Y.B.Zhang}, and edge-dependent optical selection rules \cite{Opt.Express19(2011)23350H.C.Chung, Phys.Chem.Chem.Phys.18(2016)7573H.C.Chung} and magneto-electronic properties \cite{PhysicaE42(2010)711H.C.Chung, J.Phys.Soc.Jpn.80(2011)044602H.C.Chung, Carbon109(2016)883H.C.Chung, Phys.Chem.Chem.Phys.15(2013)868H.C.Chung, Philos.Mag.94(2014)1859H.C.Chung}) stimulate the development of many low-dimensional materials and researches in various fields for the last decade \cite{Nat.Chem.5(2013)263M.Chhowalla, Adv.Funct.Mater.25(2015)5086K.Kalantar-zadeh, BookLinRichQuasiparticlePropertiesLowDimensionalSystems, Nature490(2012)192K.S.Novoselov, Adv.Mater.24(2012)210M.Osada, Chem.Rev.113(2013)3766M.S.Xu}. Up to now, low-dimensional materials with various sizes of bandgap can be synthesized (e.g., gapless graphene, insulating hexagonal boron nitride ($h$-BN) \cite{Nature579(2020)219T.A.Chen}, and semiconducting transition metal dichalcogenides (TMDs) \cite{Nat.Nanotechnol.7(2012)699Q.H.Wang} and group III-V compounds \cite{Sci.Rep.9(2019)2332H.C.Chung}), indicating that they can be building blocks similar to bulk materials. Moreover, due to their unique features (e.g., ultrathin, tunneling barrier), low-dimensional materials are expected to be used in next-generation additives in Na-ion batteries beyond the bulk materials.

The successful commercialization and popularization of EVs worldwide \cite{GlobalEVOutlook2020IEA} cause the widespread and mass production of Li-ion batteries. The retired power batteries have largely increased, causing waste of resources and environmental protection threats. Hence, recycling and utilization of such retired batteries have been promoted \cite{Sustain.EnergyTechnol.Assess.6(2014)64L.Ahmadi, Renew.Sustain.EnergyRev.93(2018)701E.Martinez-Laserna, CellRep.Phys.Sci.2(2021)100537J.Zhu}. Some retired power batteries still possess about 80\% initial capacity \cite{J.Environ.Manage.232(2019)354L.C.Casals, FMEAofLFPBatteryModule2018Chung, WorldElectr.Veh.J.9(2018)24A.Podias, J.EnergyStorage11(2017)200S.Tong, J.PowerSources196(2011)5147E.Wood}. So they can be repurposed and utilized once again, e.g., serving as the battery modules in the stationary energy storage system \cite{Batteries3(2017)10L.C.Casals, EnergyPolicy71(2014)22C.Heymans, WasteManage.113(2020)497D.Kamath, Batteries5(2019)33H.Quinard}. Governments in various countries have acknowledged this emergent issue and prepared to launch their policies to deal with the recovery and reuse of repurposing batteries, such as coding principles, traceability management system, manufacturing factory guidelines, dismantling process guidelines, residual energy measurement, federal and state tax credits, rebates, and other financial support \cite{J.TaiwanEnergy6(2019)425H.C.Chung, EnergyPolicy113(2018)535K.Gur, IEEEAccess7(2019)73215E.Hossain}.

Safety and performance are important in using retired power batteries, i.e., repurposing batteries. Underwriters Laboratories (UL), a global safety certification company established in 1894, published the standard for evaluating the safety and performance of repurposing batteries in 2018 (UL 1974) \cite{MJTE860(2020)35H.C.Chung, BookChChung2021UL1974, UL1974}. The document briefly provides a general procedure of the safety operations and performance tests on retired power battery packs, modules, and cells, while the detailed steps and specifics are not given in detail. In real-world applications, the design, form factor, and raw materials of the existing pack/modules/cells often vary greatly from one another, making it difficult to develop a unified technical procedure. Furthermore, information on the detailed technical procedures used is often not easily available in the open literature, except for Schneider et al., who reported the procedure to classification and reuse small cylindrical NiMH battery modules for mobile phones \cite{J.PowerSources189(2009)1264E.L.Schneider, J.PowerSources262(2014)1E.L.Schneider}; Zhao, who shared the successful experiences of some grid-oriented applications of EV Li-ion batteries in China in a comprehensive technical procedure \cite{ReuseRecyclingLiIonPowerBatteries2017Zhao}; and Chung, who announced the procedure described in UL 1974 on LFP repurposing batteries and released the related experimental dataset publicly \cite{Sci.Data8(2021)165H.C.Chung}. Based on the information from these studies, a conceptual procedure for repurposing applications is summarized in five primary steps: (1) judgment of the retired battery system based on historical information, (2) disassembly of retired battery packs/modules, (3) battery performance evaluation (mechanical, electrochemical, and safety), (4) sorting, and (5) developing control and management strategies for repurposing applications. In reality, multiple rounds of inspections and assessments might usually be performed \cite{Resour.Conserv.Recycl.154(2020)104461M.AlfaroAlgaba, Resour.Conserv.Recycl.159(2020)104785H.Rallo}. Overall, the technical feasibility of repurposing applications of retired EV power batteries mainly depends on whether these steps could be performed effectively and efficiently. Na-ion batteries as potential candidates for next-generation batteries should consider the possibilities in reuse and recycling of repurposing Na-ion batteries.

\section*{Acknowledgements}

The author (H. C. Chung) thanks Prof. Ming-Fa Lin for the book chapter invitation and for inspiring him to study this topic. H. C. Chung would like to thank the contributors to this article for their valuable discussions and recommendations, Jung-Feng Jack Lin, Hsiao-Wen Yang, Yen-Kai Lo, and An-De Andrew Chung. The author (H. C. Chung) thanks Pei-Ju Chien for English discussions and corrections as well as Ming-Hui Chung, Su-Ming Chen, Lien-Kuei Chien, and Mi-Lee Kao for financial support. This work was supported in part by Super Double Power Technology Co., Ltd., Taiwan, under the project “Development of Cloud-native Energy Management Systems for Medium-scale Energy Storage Systems (\href{https://osf.io/7fr9z/}{https://osf.io/7fr9z/})” (Grant number: SDP-RD-PROJ-001-2020). This work was supported in part by Ministry of Science and Technology (MOST), Taiwan (grant numbers: MOST 108-2112-M-006-016-MY3, MOST 109-2124-M-006-001, and MOST 110-2811-M-006-543).

\bibliography{Reference}
\bibliographystyle{apsrev4-2}

\end{document}